\documentclass[aps,prb,reprint,superscriptaddress]{revtex4-2}
\usepackage{amsmath,amssymb,graphicx,bm,dsfont}
\usepackage{xcolor}
\usepackage[colorlinks=true,urlcolor=blue,linkcolor=blue,citecolor=blue,bookmarks=false]{hyperref}

\usepackage{changes}

\usepackage{soul}

\graphicspath{{figures/}}

\begin{document}
	
\date{\today}
\title{Low-temperature antiferromagnetic order in orthorhombic CePdAl$_{3}$}
	
\author{Vivek Kumar}
\email{vivek.kumar@tum.de}
\affiliation{Physik-Department, Technische Universit{\"a}t M{\"u}nchen, D-85748 Garching, Germany}
	
\author{Andreas Bauer}
\affiliation{Physik-Department, Technische Universit{\"a}t M{\"u}nchen, D-85748 Garching, Germany}
\affiliation{Zentrum f{\"u}r QuantumEngineering (ZQE), Technische Universit{\"a}t M{\"u}nchen, D-85748 Garching, Germany}

\author{Christian Franz}
\affiliation{Physik-Department, Technische Universit{\"a}t M{\"u}nchen, D-85748 Garching, Germany}
\affiliation{J{\"u}lich Centre for Neutron Science (JCNS) at Heinz Maier-Leibnitz Zentrum (MLZ), D-85748 Garching, Germany}

\author{Jan Spallek}
\affiliation{Physik-Department, Technische Universit{\"a}t M{\"u}nchen, D-85748 Garching, Germany}

\author{Rudolf Sch{\"o}nmann}
\affiliation{Physik-Department, Technische Universit{\"a}t M{\"u}nchen, D-85748 Garching, Germany}

\author{Michal Stekiel}
\affiliation{Physik-Department, Technische Universit{\"a}t M{\"u}nchen, D-85748 Garching, Germany}

\author{Astrid Schneidewind}
\affiliation{J{\"u}lich Centre for Neutron Science (JCNS) at Heinz Maier-Leibnitz Zentrum (MLZ), D-85748 Garching, Germany}

\author{Marc Wilde}
\affiliation{Physik-Department, Technische Universit{\"a}t M{\"u}nchen, D-85748 Garching, Germany}
\affiliation{Zentrum f{\"u}r QuantumEngineering (ZQE), Technische Universit{\"a}t M{\"u}nchen, D-85748 Garching, Germany}
	
\author{C. Pfleiderer}
\affiliation{Physik-Department, Technische Universit{\"a}t M{\"u}nchen, D-85748 Garching, Germany}
\affiliation{Zentrum f{\"u}r QuantumEngineering (ZQE), Technische Universit{\"a}t M{\"u}nchen, D-85748 Garching, Germany}
\affiliation{Munich Center for Quantum Science and Technology (MCQST), Technische Universit{\"a}t M{\"u}nchen, D-85748 Garching, Germany}
	
	
\begin{abstract}
We report the magnetization, ac susceptibility, and specific heat of optically float-zoned single crystals of CePdAl$_{3}$. In comparison to the properties of polycrystalline CePdAl$_{3}$ reported in the literature, which displays a tetragonal crystal structure and no long-range magnetic order, our single crystals exhibit an orthorhombic structure ($Cmcm$) and order antiferromagnetically below a N\'eel temperature $T_{\rm N}$\,=\,5.6\,K. The specific heat at zero-field shows a clear $\lambda$-type anomaly with a broad shoulder at $T_{\rm N}$. A conservative estimate of the Sommerfeld coefficient of the electronic specific heat, $\gamma = 121~\mathrm{mJ~K^{-2}~mol^{-1}}$, indicates a moderately enhanced heavy-fermion ground state. A twin microstructure evolves in the family of planes spanned by the basal plane lattice vectors \textit{\textbf{a}$_{\rm o}$} and \textit{\textbf{c}$_{\rm o}$}, with the magnetic hard axis \textit{\textbf{b}$_{\rm o}$} common to all twins. The antiferromagnetic state is characterized by a strong magnetic anisotropy and a spin-flop transition induced under magnetic field along the easy direction, resulting in a complex magnetic phase diagram. Taken together our results reveal a high sensitivity of the magnetic and electronic properties of CePdAl$_{3}$ to its structural modifications. 
\end{abstract}
	
\maketitle
	
\section{Introduction}
Cerium-based intermetallic compounds exhibit a variety of ground states and various underlying exotic physical phenomena, such as unconventional superconductivity \cite{bauer2004heavy,takeuchi2004magnetism,kimura2005pressure,sugitani2006pressure,kimura2007normal,weng2016multiple,bonalde2009unusual,pfleiderer2009superconducting}, heavy-fermion states \cite{egetenmeyer2012direct,jiao2019enhancement}, non-Fermi liquid behavior \cite{stewart2001non}, vibronic hybrid excitations \cite{thalmeier1982bound,chapon2006magnetic,adroja2012vibron,klicpera2017magnetic,opagiste2011unconventional}, and complex magnetic order \cite{takayanagi1994two,nakotte1994complex,thamizhavel2007anisotropic,nakano2019coexistence,settai1997single,das1997transport,pecharsky1993unusual,pikul2003kondo}. On the phenomenological level, the origin of this remarkable diversity of ground states has been attributed to the competition of narrow f-electron bands and strong electronic correlations together with spin-orbit interaction, crystal electric field (CEF) effects, and strong magneto-elastic coupling. An overarching theme connecting much of the research in f-electron compounds concerns the condition of the emergence of magnetic order. 

A class of compounds with the general formula Ce\textit{T}\textit{X}$_{3}$ (\textit{T} is a transition metal and \textit{X} is a $p$-block element) crystallizing in subgroups of the BaAl$_{4}$-type ($I4/mmm$) tetragonal structure has received special attention \cite{mentink1993antiferromagnetism,lee1994competition,moze1996crystal,paschen1998transport,klicpera2014structural,klicpera2015magnetization,kontani1994magnetic,mock1999low,hillier2012muon,muranaka2007thermodynamic,nallamuthu2017ferromagnetism,mock1999low,das1997transport,pikul2003kondo,pecharsky1993unusual,thamizhavel2005unique,kaneko2009multi,smidman2013neutron,kawai2007magnetic,kimura2005pressure,weng2016multiple,wang2019anomalous,terashima2007fermi,kawai2007pressure}. In these compounds, a large number of structural variants and diverse magnetic and electrical properties can be obtained by changing the transition metal \textit{T}. Many members of this class such as CeRhGe$_{3}$, CeAuAl$_{3}$, CeCuAl$_{3}$, and CeCoGe$_{3}$ adopt a non-centrosymmetric tetragonal structure (BaNiSn$_{3}$-type $I4mm$) and exhibit antiferromagnetic behavior \cite{mentink1993antiferromagnetism,lee1994competition,moze1996crystal,paschen1998transport,klicpera2014structural,klicpera2015magnetization,kontani1994magnetic,mock1999low,hillier2012muon}. Other members such as CeAgAl$_{3}$ display ferromagnetism with a centrosymmetric orthorhombic crystal structure \cite{muranaka2007thermodynamic,nallamuthu2017ferromagnetism}. A spin-glass state was reported in non-centrosymmetric tetragonal CePtAl$_{3}$ below 0.8\,K \cite{mock1999low}. Complex magnetic phases have been observed in antiferromagnetic CeNiGe$_{3}$\cite {das1997transport,pikul2003kondo}, CeCoGe$_{3}$ \cite{pecharsky1993unusual,thamizhavel2005unique,kaneko2009multi,smidman2013neutron} and CePtSi$_{3}$ \cite{kawai2007magnetic}. The discovery of pressure-induced unconventional superconductivity in the non-centrosymmetric tetragonal heavy-fermion antiferromagnets CeRhSi$_{3}$, CeIrSi$_{3}$, CeCoGe$_{3}$, CeIrGe$_{3}$, and CeRhGe$_{3}$ even suggests a new direction in condensed matter physics \cite{kimura2005pressure,weng2016multiple,wang2019anomalous,terashima2007fermi,kawai2007pressure}.

An important aspect is the structural stability of these systems and the emergence of different electronic ground states. As one of the first examples,  CePd$_{2}$Al$_{2}$ \cite{chapon2006magnetic,klicpera2017magnetic}, which is closely related to the class of Ce\textit{T}Al$_{3}$ of materials, was found to undergo a structural phase transformation from a tetragonal to an orthorhombic lattice at 13.5\,K. An inelastic neutron scattering study revealed three magnetic excitations in the paramagnetic phase. However,  according to Kramer’s theorem, only two CEF excitations are expected due to the splitting of ground state \textit{J}\,=\,5/2 of the Ce$^{3+}$ ion into three doublets in tetragonal/orthorhombic point symmetry suggesting strong coupling between the crystal fields and the crystal structure. Later, Adroja \emph{et al.} found a similar anomaly in CeCuAl$_{3}$ \cite{adroja2012vibron}, where a structural instability manifests itself in terms of a drastic change in lattice parameters of the tetragonal structure around 300\,$^{\circ}$C \cite{klicpera2014structural}. These anomalous excitations have been interpreted by means of Thalmeier and Fulde's model of bound states between phonons and CEF excitations as generalized to the tetragonal point symmetry.  Recently, \v{C}erm\'{a}k \emph{et al.} confirmed related hybrid CEF-phonon excitations even for weak magnetoelastic coupling in isostructural CeAuAl$_{3}$ \cite{vcermak2019magnetoelastic}. Moreover, CePd$_{2}$Al$_{2}$, CeCuAl$_{3}$ and CeAuAl$_{3}$ order antiferromagnetically at low temperatures and exhibit incommensurate amplitude-modulated magnetic structures \cite{klicpera2015neutron,matsumura200927al,adroja2015muon,klicpera2017magnetic}. The presence of multi-step magnetism and complex magnetic phase diagrams suggests the possible existence of topologically non-trivial multi-k structures akin to skyrmion lattices \cite{muhlbauer2009skyrmion}. This raises the question, if and how the formation of magnetic order depends on the stabilization of specific crystal structure.

In this paper we focus on CePdAl$_{3}$. A study of as-cast polycrystalline CePdAl$_{3}$ by Schank \emph{et al.} in 1994 revealed a tetragonal $I4mm$ structure with lattice constants \textit{a} = 4.343\,Å and \textit{c} = 10.578\,Å \cite{schank19944f}, where the heat treatment at high temperature results in a structural phase transformation with an antiferromagnetic order below $T_{\rm N}$ $\simeq$ 6\,K. In contrast, no magnetic order was found down to 0.1\,K in a recent investigation by Franz \emph{et al.} on single crystalline tetragonal CePdAl$_{3}$ grown by optical float zoning with a growth rate of 6\,mm/h \cite{franz2016single}. For the work reported in the following, a single crystal was prepared by optical float zoning using a much lower growth rate of 1\,mm/h. Under these conditions we found that CePdAl$_{3}$ crystallizes in an orthorhombic as opposed to a tetragonal structure \cite{Rudolf:Thesis:2015}. In this paper, we report comprehensive magnetization, ac susceptibility, and specific heat measurements on single crystalline orthorhombic CePdAl$_{3}$. As our main result we find the characteristics of antiferromagnetic order below $T_{\rm N}$\,=\,5.6\,K. We determine the magnetic phase diagram upto 14\,T, where we find the emergence of complex magnetic phases under magnetic fields applied along the easy direction. The presence of different structural and magnetic configurations of CePdAl$_{3}$ identifies a new example of a material in which to search for hybrid excitations and new magnetic phases in the future.
  
Our paper is organized as follows. After a brief account of the experimental methods in Sec.\,II, we present our experimental results in Sec.\,III. We start with the structural properties and notation in Sec.\,III\,A, followed by the specific heat results in Sec.\,III\,B and magnetic susceptibility data in Sec.\,III\,C. The temperature- and field-dependence of the magnetization is presented in Sec.\,III\,D. We find that the magnetic field-driven transitions for fields applied along the easy direction are consistent with the specific heat as a function of temperature as presented in Sec.\,III\,E. In Sec.\,III\,F, we examine the magnetic transitions in more detail by analyzing the hysteresis of the field-dependent magnetic susceptibility. Comprehensive datasets allow to infer the magnetic phase diagram presented in Sec.\,III\,G. The conclusions are summarized in Sec.\,IV.
	
\section{Experimental methods}

A single-crystal of CePdAl$_{3}$ was grown using the optical floating-zone technique following a  process similar that described in Ref.\,\cite{franz2016single,neubauer2011ultra,bauer2016ultra}. As the main difference, the growth rate was reduced from 6\,mm/h \cite{franz2016single} to 1\,mm/h which resulted in the formation of an orthorhombic crystal.

The crystal structure of CePdAl$_{3}$ was determined by means of single-crystal x-ray diffraction (SCXRD). A platelet-shaped crystal with dimensions 50\,$\mu$m\,$\times$\,40\,$\mu$m\,$\times$ 10\,$\mu$m was cleaved of the CePdAl$_{3}$ crystal as grown.  The platelet was investigated at a Rigaku XtaLAB Synergy-S diffractometer, using a Mo\,x-ray source with $\lambda$ = 0.71\,Å and a two-dimensional HyPix-Arc 150$^{\circ}$ detector. Bragg reflections were indexed using CrysAlis$^{Pro}$ \cite{crysalispro2014agilent} as integrated with the diffractometer.  

The single crystals were oriented by Laue\,x-ray diffraction and a cuboidal sample was cut with orientations \textit{\textbf{a}$_{\rm o}^{\star}$}, \textit{\textbf{c}$_{\rm o}^{\star}$} and \textit{\textbf{b}$_{\rm o}$} as introduced below for the measurement of the bulk properties. The ac susceptibility, magnetization, and specific heat were measured in a  Quantum Design physical property measurement system (PPMS) at temperatures down to 2\,K under magnetic fields up to 14\,T. In order to determine the temperature dependence of the bulk properties, the sample was first cooled from a high temperature, well above $T_{\rm N}$, to the lowest attainable temperature in the absence of a magnetic field. Subsequently, the field was set to the desired value and data were collected for increasing temperature. This protocol was repeated for different target magnetic fields. The ac susceptibility was measured at an excitation amplitude of 1\,mT and an excitation frequency of 911\,Hz. The specific heat was measured down to 2\,K using a large heat-pulse method \cite{bauer2013specific}. For temperatures between 0.08\,K and 4\,K the specific heat was measured in a Dryogenic adiabatic demagnetization refrigerator using a conventional heat-pulse method.

The field dependence of the magnetization and the ac susceptibility was measured using the following temperature versus field protocol. First, the sample was cooled from a high temperature well above $T_{\rm N}$ to the target temperature in the absence of a magnetic field. Second, data as a function of magnetic field were recorded in a sequence of field sweeps from zero-field to 14\,T, 14\,T to -14\,T, and -14\,T to 14\,T.

The bulk properties recorded on different pieces cut from the large single crystal ingot were consistent. The temperature and field dependent features along \textit{\textbf{a}$_{\rm o}^{\star}$} and \textit{\textbf{c}$_{\rm o}^{\star}$} were qualitatively identical. Therefore, comprehensive data focused on one of these directions, \textit{\textbf{c}$_{\rm o}^{\star}$}, were recorded. Summarizing the key result of our study, the magnetic phase diagrams of CePdAl$_3$ were inferred. Signatures detected in measurements as a function of temperature and magnetic field are labelled as $T_j$ and $H_j$, respectively. For clarity, the same subscript $j$ is assigned to the transitions corresponding to the same line in the phase diagram.

\section{Experimental results}

\subsection{Crystal structure and twinning}

Different crystal growth conditions favor a tetragonal ($I4mm$) \cite{franz2016single} or orthorhombic crystal structures of CePdAl$_3$. By means of single crystal x-ray diffraction, we determined that the orthorhombic lattice stabilizes in the $Cmcm$ space group. The lattice parameters at room temperature are $a_o$ = 6.379\,Å, $b_o$ = 10.407\,Å and $c_o$ = 5.975\,Å. The orthorhombic phase exhibits a pseudo-tetragonal twinning in the basal plane, evident, for instance, by the splitting of the Bragg reflections shown in Fig.\,1(a). The twinning law was determined by indexing all measured reflections with components of the four twins presented in Fig.\,1(b). An illustration of the twin orientation is shown in Figs.\,1(c) and (d). The three perpendicular cartesian directions of twins for $i=1,2,3,4$ are denoted by \textit{\textbf{a}$_{\rm o}^i$}, \textit{\textbf{b}$_{\rm o}^i$} and \textit{\textbf{c}$_{\rm o}^i$}, where \textit{\textbf{a}$_{\rm o}^i$} and \textit{\textbf{c}$_{\rm o}^i$} construct an effective basal plane and \textit{\textbf{b}$_{\rm o}^i$} mutually represents the long axis. The volume fraction of the four twins labelled $i$ = 1, 2, 3, and 4 are 0.38, 0.26, 0.23, and 0.13, respectively. The mismatch angle between the twins numbered 1 and 2, as well as 3 and 4, are around 3$^{\circ}$. 

Measurements on different pieces cleaved of the single crystalline ingot demonstrate the same twinning scheme with minor differences in twin fractions of different twins. An attempt to detwin the crystals by means of high-temperature treatment, etching, or cleaving of micrometer-sized crystals neither affected the twinning as such nor the twinning fractions.

In turn, measurements in any direction in the effective basal plane reflect effectively an admixture of \textit{\textbf{a}$_{\rm o}^i$} and \textit{\textbf{c}$_{\rm o}^i$} directions due to the four twins.  We define, therefore, two mutually perpendicular effective \emph{sample} directions \textit{\textbf{a}$_{\rm o}^{\star}$} and \textit{\textbf{c}$_{\rm o}^{\star}$}, explicitly taking into account the volume fractions of the four twins. This definition is schematically depicted in Figs.\,1(c) and (d) where \textit{\textbf{a}$_{\rm o}^{\star}$} is nearly aligned along \textit{\textbf{a}$_{\rm o}^{1,2}$} and \textit{\textbf{c}$_{\rm o}^{3,4}$}, while \textit{\textbf{c}$_{\rm o}^{\star}$} is aligned to that of \textit{\textbf{c}$_{\rm o}^{1,2}$} and \textit{\textbf{a}$_{\rm o}^{3,4}$}.  The third crystal direction, corresponding to the long axis \textit{\textbf{b}$_{\rm o}$}, remains unaffected by the twin deformations.

\begin{figure}
	\includegraphics[width=1.0\linewidth]{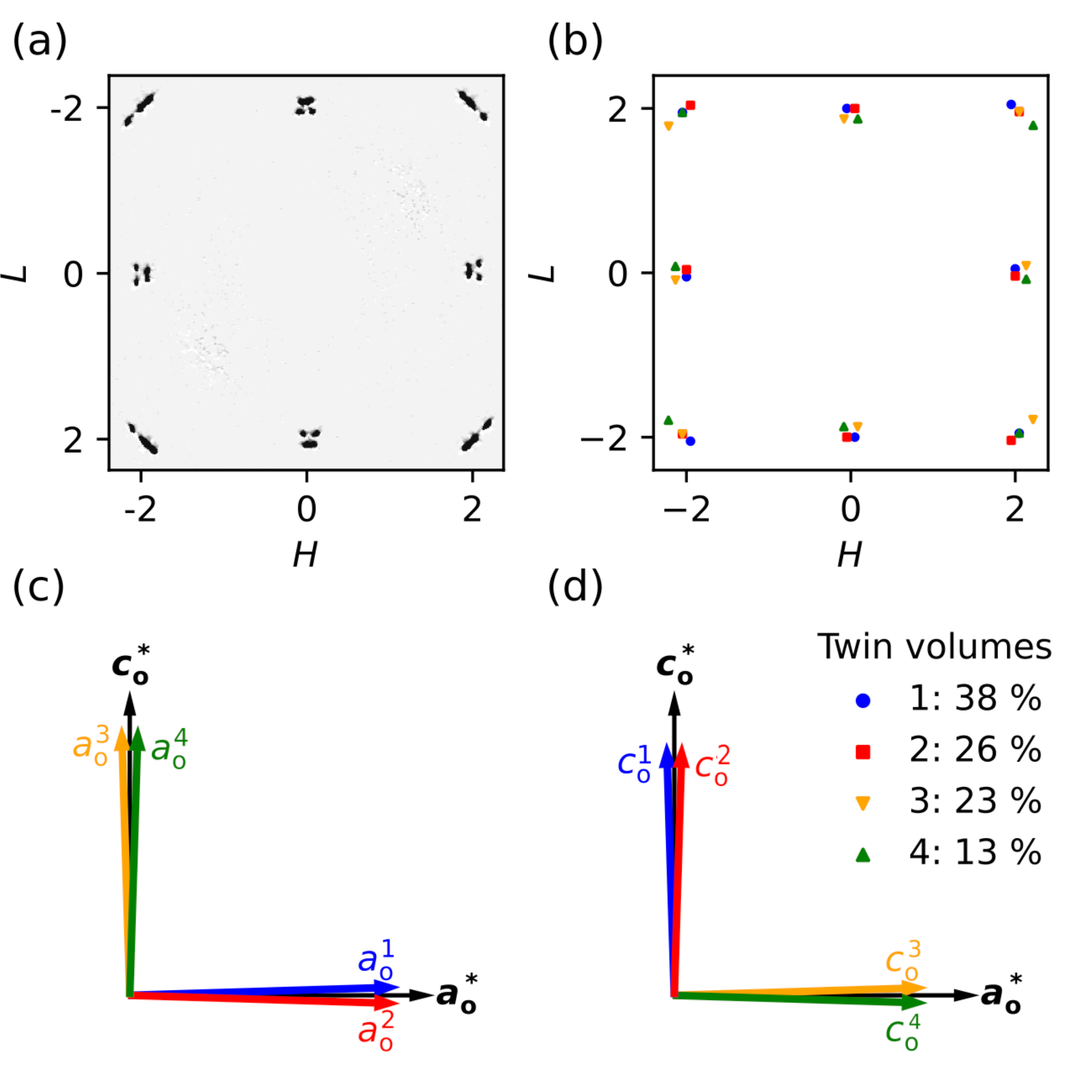}
	\caption{\label{figure1}Twin scheme in the basal plane of orthorhombic CePdAl$_3$ as derived from single crystal x-ray diffraction. (a) X-ray scattering intensity reconstructed in the $H0L$ plane. The splitting of the reflections is characteristic of twin formation. (b) Indexed reflections of panel (a) with the colors corresponding to different twin domains. Schematics of the lattice vectors  \textit{\textbf{a}$_{\rm o}^i$} and \textit{\textbf{c}$_{\rm o}^i$} of twin $i$ in the basal plane $H0L$ of the orthorhombic crystal are depicted in the lower panels (c) and (d). Four twins labelled $i$ = 1, 2, 3, and 4 were identified. \textit{\textbf{a}$_{\rm o}^{\star}$} and \textit{\textbf{c}$_{\rm o}^{\star}$} are defined as mutual perpendicular sample directions comprising the admixtures of twin lattice vectors.} 
\end{figure}
	
\subsection{Temperature-dependence of the specific heat}
The temperature dependence of the specific heat \textit{C}(\textit{T}) of single-crystalline tetragonal ($I4mm$) and orthorhombic ($Cmcm$) CePdAl$_{3}$, as well as nonmagnetic polycrystalline tetragonal ($I4mm$) LaPdAl$_{3}$ measured in the absence of a magnetic field are shown in Fig.\,2. No evidence suggesting magnetic order was observed in tetragonal CePdAl$_{3}$ \cite{franz2016single}.  In orthorhombic CePdAl$_{3}$, a $\lambda$-type anomaly comprising a peak at 5.4\,K followed by a shoulder closely above the transition temperature $T_{\rm N}$\,=\,5.6\,K is observed, where the magnetization is characteristic of antiferromagnetism as reported below. The behavior observed is consistent with a previous study of polycrystalline CePdAl$_{3}$ \cite{schank19944f}. Moreover, the properties are reminiscent of the commensurate to incommensurate magnetic transition reported of other strongly correlated systems \cite{mishra2021specific,kumar2010magnetic}. 

A pronounced shoulder in the specific heat has also been seen in other systems, notably, the chiral cubic magnet MnSi \cite{stishov2007magnetic,bauer2013specific}, where it reflects a change of character of the critical spin-fluctuations when approaching long-range helimagnetic order and a concomitant fluctuation-induced first-order transition. Details of the low-temperature specific heat of orthorhombic CePdAl$_{3}$ at zero-field are presented in Sec.\,III\,E below, which also includes data collected at different magnetic fields.

Above $T_{\rm N}$, the expression ${C}\slash{T}=\gamma+\beta{T}^{2}$, where $\gamma$ and $\beta$ are the electronic and phononic contributions to the specific heat, respectively, has been fitted to the specific heat data in the range $\sim$18 to $\sim$23 K of orthorhombic CePdAl$_{3}$. The values obtained for $\gamma$ and $\beta$ are 234 $\mathrm{mJ~mol^{-1}~K^{-2}}$ and 3.437 $\times$ $10^{-4}$ $\mathrm{J~mol^{-1}~K^{-4}}$, respectively. The Debye temperature, $\Theta_{\rm D}$ = 305 K, associated with $\beta$ may be derived using the relation $\beta = (12/5)\pi^{4}n{\rm R}\slash\Theta_{\rm D}^{3}$, where \textit{n} is the number of atoms per formula unit and R is the gas constant. The phonon contribution to the specific heat in the Debye model [orange line in Fig.\,2] is given by
\begin{equation}\label{equation1}
\begin{split}
C_{\rm ph,Debye} = &~9n{\rm R}\left(\frac{T}{\Theta_{\rm D}}\right)^{3}\int_{0}^{x_{\rm D}}{\frac{x^{4}e^x}{(e^x-1)^2}}dx
\end{split}
\end{equation}
where $x_{\rm D}=\Theta_{\rm D}/T$. At high temperatures the experimental data of tetragonal LaPdAl$_{3}$ and CePdAl$_{3}$, as well as orthorhombic CePdAl$_{3}$  approach the Dulong-Petit limit, 3\textit{n}R = 15R = 124.7 $\mathrm{J~mol^{-1}~K^{-1}}$, where \textit{n} = 5.

The large value of $\gamma$ = 234\,$\mathrm{mJ~mol^{-1}~K^{-2}}$ obtained from the low-temperature specific heat above $T_{\rm N}$ is typical for a heavy-fermion system. It has to be borne in mind, however, that evaluating $\gamma$ at the relatively high-temperature range above $T_{\rm N}$ is associated with substantial uncertainties. A lower bound of $\gamma$, fitting the experimental data in the antiferromagnetic state at temperatures between $\sim$0.9\,K and $\sim$3.7\,K, yields a value of $\gamma$ = 121\,$\mathrm{mJ~mol^{-1}~K^{-2}}$ still characteristic of heavy-fermion behaviour.

At high temperatures ($T$ $>$ 100\,K), the specific heat of all three compounds exhibits essentially the same temperature dependence. However, the specific heat of orthorhombic CePdAl$_{3}$ is slightly smaller than for tetragonal CePdAl$_{3}$, suggesting reduced electronic and phononic contributions associated with the reduced crystal symmetry. Compared to nonmagnetic LaPdAl$_{3}$, the specific heat of orthorhombic CePdAl$_{3}$ is also slightly smaller, yet within the experimental error of experiment. Indeed, a multiplication with a fraction of 0.99 to the total signal of LaPdAl$_{3}$ fully superimposes the data of CePdAl$_{3}$ as shown in Fig.\,3(a) of $C/T$ vs $T$. The corresponding difference in specific heats may be attributed to the magnetic contribution of the specific heat of orthorhombic CePdAl$_{3}$.

Shown in Fig.\,3(b) is a sharp peak at $T$\,=\,5.4\,K in the magnetic contribution to the specific heat following subtraction of the phonon contribution signaling an antiferromagnetic transition. In addition, a broad maximum around 30\,K may be discerned as characteristic of a Schottky anomaly due to crystal electric field contributions. 

In the tetragonal as well as the orthorhombic symmetry of the lattice, the degeneracy of the sixfold ground state multiplet of the Ce$^{3+}$ ion splits into three doublet states. These lift the first and second excited state with respect to the ground state resulting in a contribution to the specific heat which can be expressed as \cite{lethuillier1976sign}
\begin{equation}\label{equation2}
\begin{split}
C_{\rm CEF} = &\frac{\rm R}{Z} \sum_{l=0}^{2} g_l\left(\frac{E_l}{kT}\right)^2 {\rm exp}\left(-\frac{E_l}{kT}\right)  \\* 
& -\frac{\rm R}{Z^2}\left(\sum_{l=0}^{2}\frac{E_l}{kT}g_l{\rm exp}\left(-\frac{E_l}{kT}\right)\right)^2 
\end{split}
\end{equation}
where 
\begin{equation}\label{equation3}
\begin{split}
Z =  \sum_{l=0}^{2} g_l{\rm exp}\left(-\frac{E_l}{kT}\right)
\end{split}
\end{equation}
is the partition function, and $l$ = 0, 1 and 2 denote the ground, first and second excited states, respectively. The degeneracy of the three doublet states is $g_0=g_1=g_2=2$. 

The energy difference $E_1-E_0 = \Delta_1$ and $E_2-E_0 = \Delta_2$ represent the levels of the first and the second excited states, respectively. A fit of the data to Eqn.\,(2) between 20\,K and 100\,K yields $\Delta_1$ = 25.4\,K and $\Delta_2$ = 76.0\,K, respectively. Note that, the normalized subtraction of the LaPdAl$_{3}$ signal may introduce systematic errors in the determination of the precise values of the excited states. For instance, subtraction of the signal of LaPdAl$_{3}$ after multiplication with a fraction of 0.98 yields $\Delta_1$ = 28.6\,K and $\Delta_2$ = 95.5\,K.

Furthermore, we have calculated the magnetic entropy $S = \int{(C/T})dT$ presented in Fig.\,3(c). At the magnetic transition temperature, the entropy reaches the theoretical value of R$\ln{2}$ for a doublet ground state expected of Ce$^{3+}$ ions. When increasing the temperature, the entropy increases and reaches Rln4 around 30\,K, approaching saturation above 100\,K consistent with the scheme of crystal electric field levels.

\begin{figure}
	\includegraphics[width=1.0\linewidth]{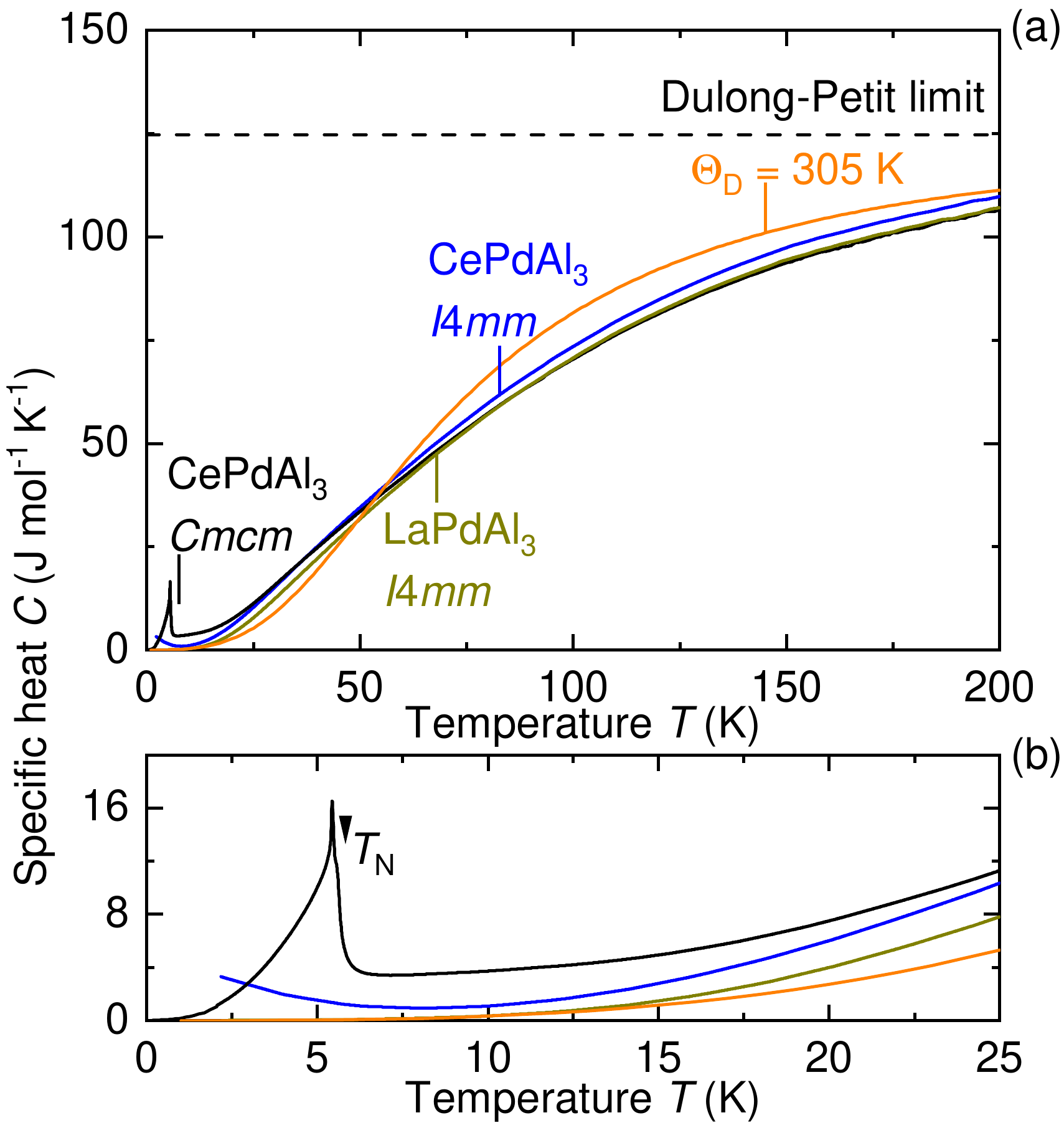}
	\caption{\label{figure2} (a) Zero-field specific heat of single-crystalline orthorhombic (black) and tetragonal (blue) \cite{franz2016single} CePdAl$_{3}$ as a function of temperature. Data of orthorhombic CePdAl$_{3}$ were measured in a Dryogenic system between 0.08\,K and 4\,K, and in a PPMS between 2\,K and 200\,K. Also shown are the specific heat of nonmagnetic polycrystalline LaPdAl$_{3}$ (Gray line) and the Debye fit (orange line) calculated from the low-temperature specific heat of the $Cmcm$ structure. The Debye temperature is $\Theta_{\rm D}$ = 305\,K. The Dulong-Petit limit for all three compounds, 15R = 124.7\,$\mathrm{J~mol^{-1}~K^{-1}}$  is depicted by a dashed line. (b) The low-temperature part of the specific heat of orthorhombic CePdAl$_{3}$ shows a pronounced $\lambda$-type anomaly with a broad shoulder at the magnetic transition at $T_{\rm N}$.} 
\end{figure}

\begin{figure}
	\includegraphics[width=1.0\linewidth]{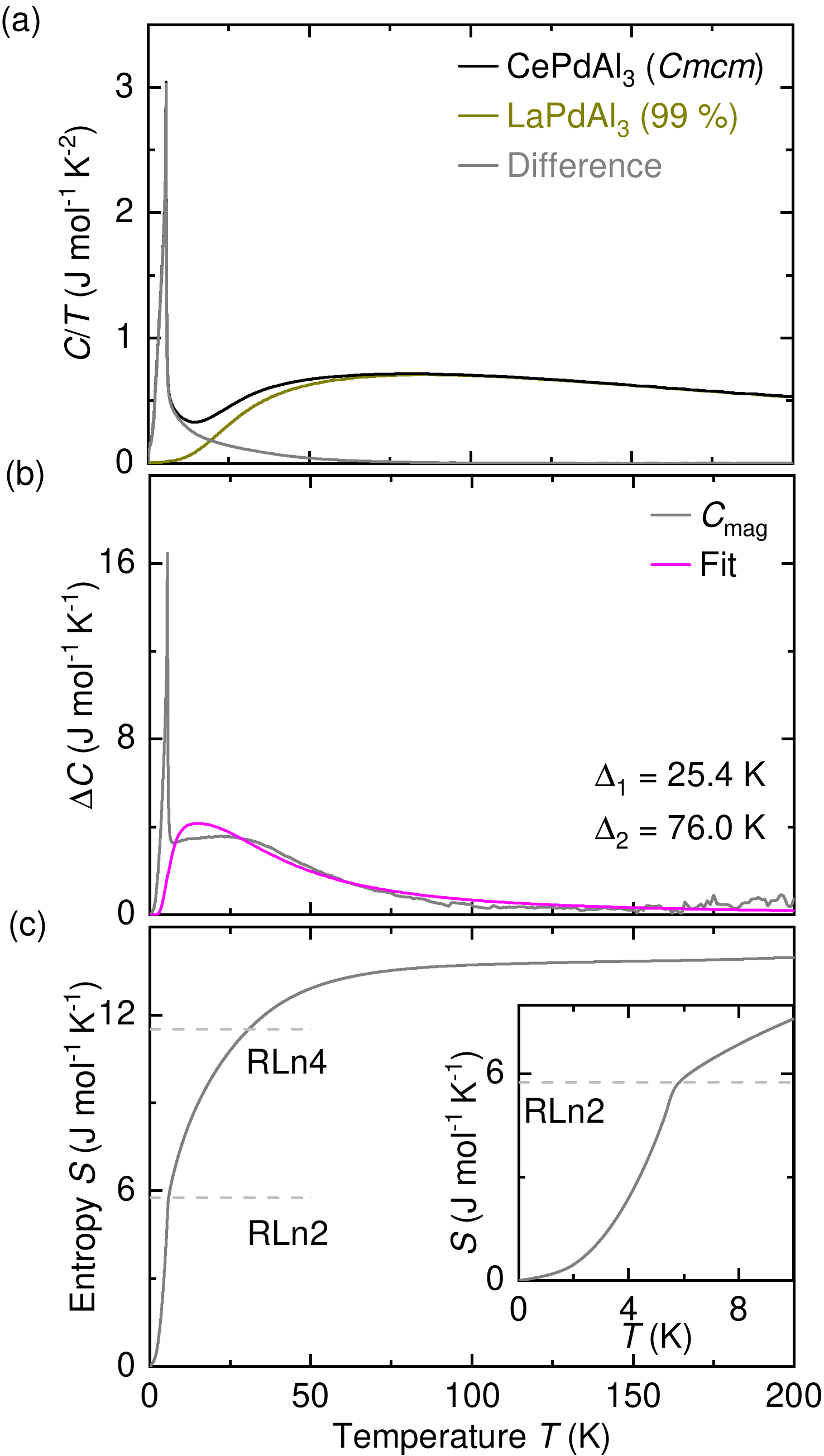}
	\caption{\label{figure3}Magnetic contribution to the specific heat and crystal electric field levels. (a) Specific heat per unit temperature, $C/T$, of orthorhombic CePdAl$_{3}$ and tetragonal LaPdAl$_{3}$ (with a multiplication of a fraction of 0.99) as well as their difference. (b) Magnetic specific heat, $C_{\rm mag}$, and the fit to the expression for the crystal electric field contribution to the specific heat yields $\Delta_1$ = 25.4\,K and $\Delta_2$ = 76.0\,K (c) Magnetic contribution to the entropy. The inset shows the entropy at low temperatures.} 
\end{figure}

\subsection{Temperature-dependence of the magnetic susceptibility}

\begin{figure}
	\includegraphics[width=1.0\linewidth]{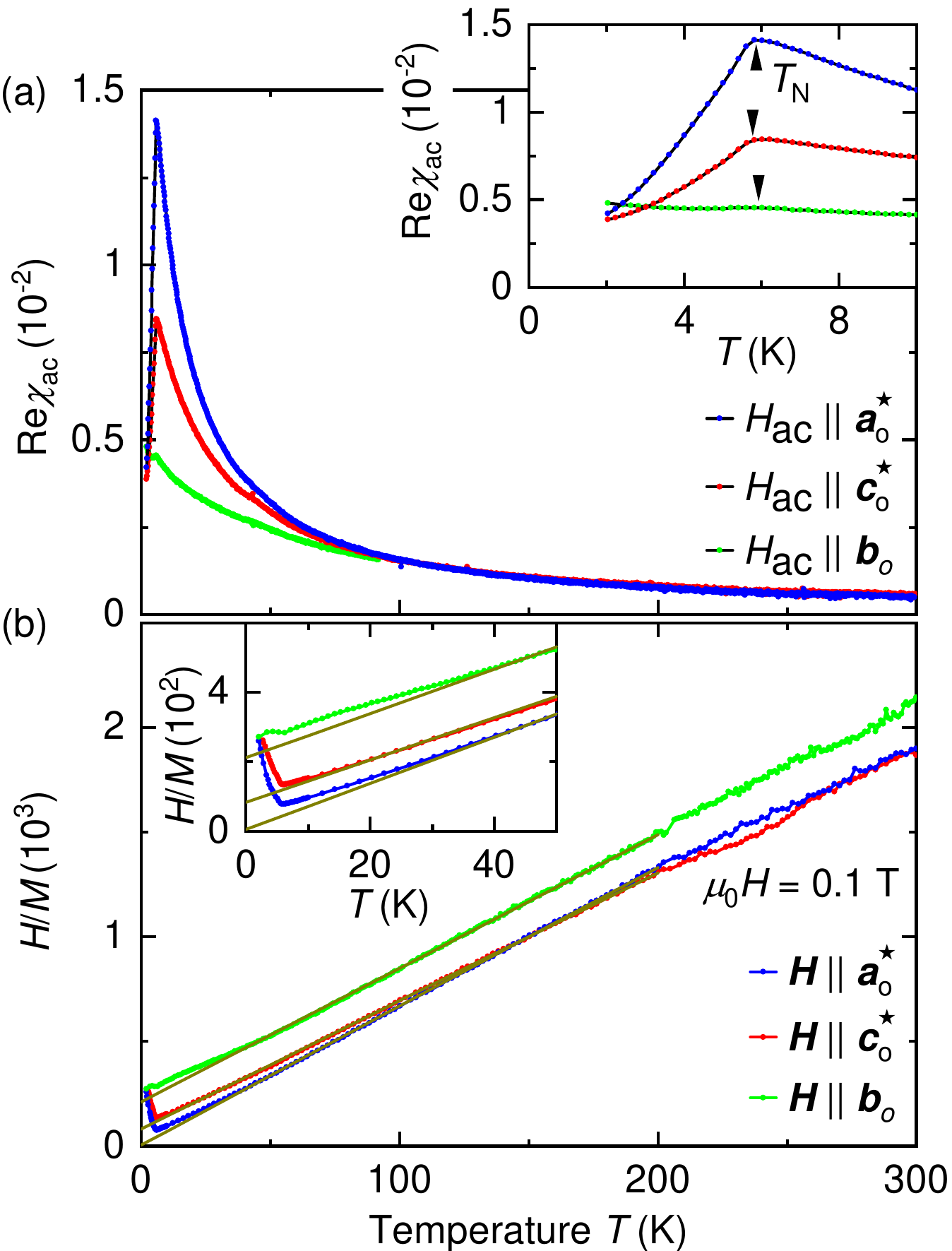}
	\caption{\label{figure4}(a) Temperature dependence of the real part of the ac susceptibility, Re\,$\chi$$_{\rm ac}$ of orthorhombic CePdAl$_{3}$ measured along \textit{\textbf{a}$_{\rm o}^{\star}$}, \textit{\textbf{c}$_{\rm o}^{\star}$} and \textit{\textbf{b}$_{\rm o}$} at an excitation amplitude of 1\,mT and a frequency of 911\,Hz. The inset shows the low-temperature part of Re\,$\chi$\,$_{\rm ac}$, reflecting the characteristics of an antiferromagnetic transition at $T_{\rm N}$\,=\,5.6\,K. (b) Susceptibility, \textit{H}\slash\textit{M}, as a function of temperature for \textit{\textbf{H}}  $\parallel$ \textit{\textbf{a}$_{\rm o}^{\star}$}, \textit{\textbf{H}}  $\parallel$ \textit{\textbf{c}$_{\rm o}^{\star}$} and \textit{\textbf{H}} $\parallel$ \textit{\textbf{b}$_{\rm o}$} measured in a field of 0.1\,T. Gray lines are Curie-Weiss fits. The inset shows the data for temperatures below 50\,K.} 
\end{figure}

The real part of the ac susceptibility, Re\,$\chi$$_{\rm ac}$ of orthorhombic CePdAl$_{3}$ as a function of temperature is shown in Fig.\,4(a) for \textit{\textbf{a}$_{\rm o}^{\star}$}, \textit{\textbf{c}$_{\rm o}^{\star}$} and \textit{\textbf{b}$_{\rm o}$}. A clear magnetic transition is observed at $T_{\rm N}$\,=\,5.6\,K in the low temperature range, characteristic of the onset of antiferromagnetic order as indicated by arrows in the inset. Namely, below $T_{\rm N}$, Re\,$\chi$$_{ac}$ monotonically decreases along \textit{\textbf{a}$_{\rm o}^{\star}$} and \textit{\textbf{c}$_{\rm o}^{\star}$} with decreasing temperature, while slightly increasing along \textit{\textbf{b}$_{\rm o}$}. The magnitude of R\,e$\chi$$_{\rm ac}$ along different axes differs significantly for \textit{T} \textless  100\,K, indicating sizeable magnetic anisotropy. 

Figure\,4(b) shows the normalized susceptibility, \textit{H}\slash\textit{M}, as a function of temperature in a field of 0.1\,T for \textit{\textbf{H}}  $\parallel$ \textit{\textbf{a}$_{\rm o}^{\star}$}, \textit{\textbf{H}}  $\parallel$ \textit{\textbf{c}$_{\rm o}^{\star}$} and \textit{\textbf{H}}  $\parallel$ \textit{\textbf{b}$_{\rm o}$}. In the paramagnetic state well above $T_{\rm N}$, a Curie-Weiss dependence is observed. A linear fit to the data above 100\,K yields Weiss temperatures $\Theta_{\rm W}^{a^{\star}}$\,=\,-0.8\,K, $\Theta_{\rm W}^{c^{\star}}$\,=\,-13.5\,K and $\Theta_{\rm W}^{b}$\,=\,-33.0\,K for \textit{\textbf{H}}  $\parallel$ \textit{\textbf{a}$_{\rm o}^{\star}$}, \textit{\textbf{H}}  $\parallel$ \textit{\textbf{c}$_{\rm o}^{\star}$} and \textit{\textbf{H}}  $\parallel$ \textit{\textbf{b}$_{\rm o}$}, respectively, characteristic of an antiferromagnetic coupling. Moreover, the effective moments of 2.39, 2.49 and 2.44 $\mu$$_{\rm B}$ per ion obtained under magnetic field along \textit{\textbf{a}$_{\rm o}^{\star}$}, \textit{\textbf{c}$_{\rm o}^{\star}$} and \textit{\textbf{b}$_{\rm o}$}, respectively, are close to the value of 2.54\,$\mu$$_{\rm B}$ expected for a free Ce$^{3+}$ ion. This might suggest a localized nature of the Ce moments in CePdAl$_{3}$. The deviation of \textit{H}\slash\textit{M} from the Curie-Weiss dependence for \textit{T$_{\rm N}$}\,\textless\,\textit{T}\,\textless\,100\,K shown in the inset of Fig.\,4(b) may be related to CEF effects and electronic correlations. Furthermore, despite the twin deformations, a significant difference between the susceptibilities along \textit{\textbf{a}$_{\rm o}^{\star}$} and \textit{\textbf{c}$_{\rm o}^{\star}$} in the paramagnetic state indicates a large anisotropy in the basal plane, characteristic of an easy-axis system.

\subsection{Magnetization}

The magnetic field dependence of the isothermal magnetization of orthorhombic CePdAl$_{3}$ at 2\,K for \textit{\textbf{H}}  $\parallel$ \textit{\textbf{a}$_{\rm o}^{\star}$}, \textit{\textbf{H}}  $\parallel$ \textit{\textbf{c}$_{\rm o}^{\star}$} and \textit{\textbf{H}}  $\parallel$ \textit{\textbf{b}$_{\rm o}$} is shown in Fig.\,5(a). No hysteresis is observed. The magnetization varies linearly in the low-field region up to 1\,T as shown in the inset of Fig.\,5(a) consistent with antiferromagnetic order. For fields along \textit{\textbf{a}$_{\rm o}^{\star}$} and \textit{\textbf{c}$_{\rm o}^{\star}$}, an $S$-shaped rise is observed in the magnetization when further increasing field. A kink around 5.5\,T suggests a field-driven transition. The magnetization values at 5.5\,T are 0.85\,$\mu$$_{\rm B}$ for \textit{\textbf{H}} $\parallel$ \textit{\textbf{a}$_{\rm o}^{\star}$}, 0.44\,$\mu$$_{\rm B}$ for \textit{\textbf{H}}  $\parallel$ \textit{\textbf{c}$_{\rm o}^{\star}$}, and 0.18\,$\mu$$_{\rm B}$ for \textit{\textbf{H}}  $\parallel$  \textit{\textbf{b}$_{\rm o}$}. The magnetization increases monotonically above this transition where the moments along \textit{\textbf{a}$_{\rm o}^{\star}$} and \textit{\textbf{c}$_{\rm o}^{\star}$} at 14\,T, the highest field strength of studied, are 1.3 and 0.7\,$\mu$$_{\rm B}$ per Ce atom, respectively. In comparison, the magnetization increases linearly with field for \textit{\textbf{H}}  $\parallel$ \textit{\textbf{b}$_{\rm o}$}. The moment at 14\,T is 0.4\,$\mu$$_{\rm B}$ per Ce atom.

\begin{figure}
	\includegraphics[width=1.0\linewidth]{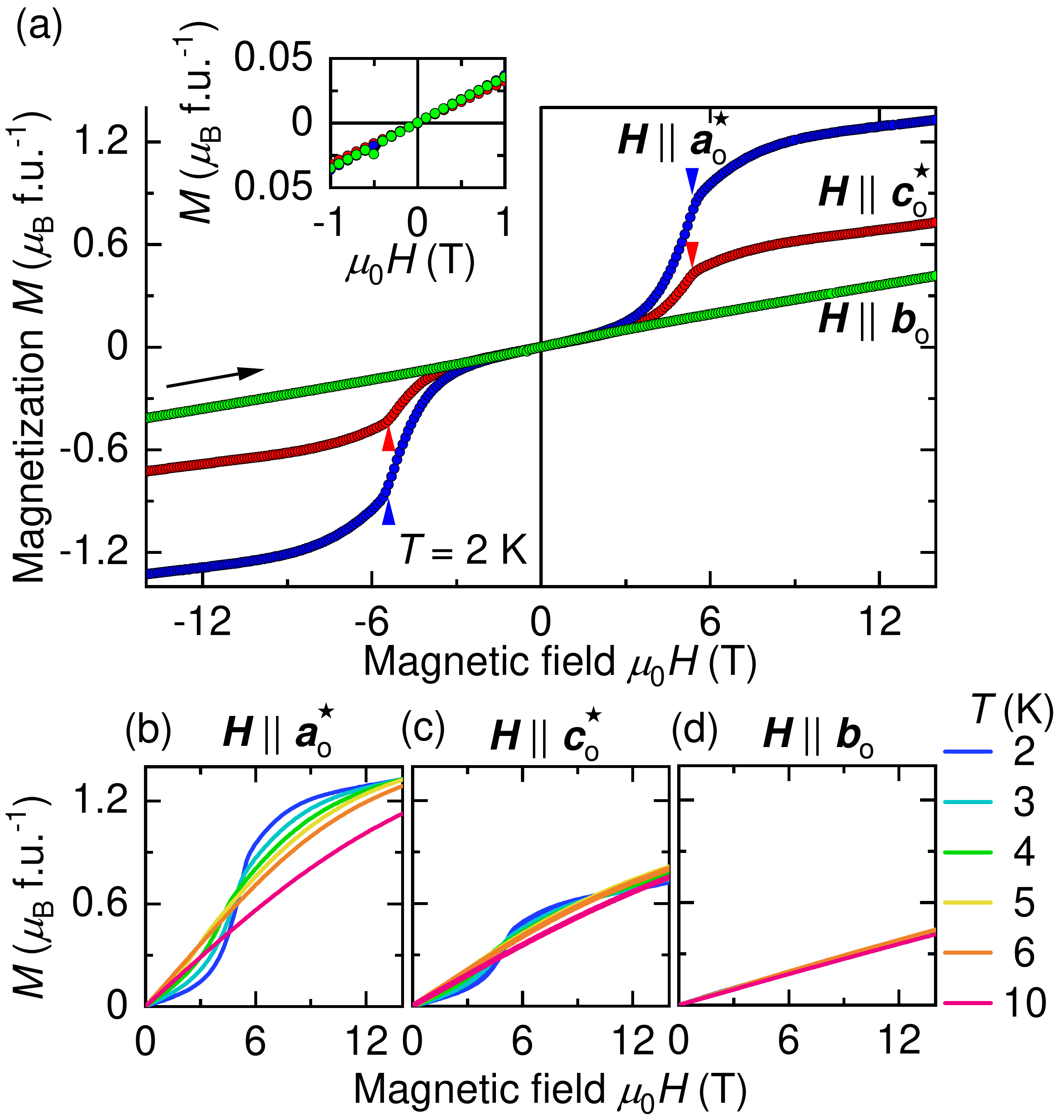}
	\caption{\label{figure5}(a) Isothermal magnetization of orthorhombic CePdAl$_{3}$ at 2\,K measured in a field along \textit{\textbf{a}$_{\rm o}^{\star}$}, \textit{\textbf{c}$_{\rm o}^{\star}$} and \textit{\textbf{b}$_{\rm o}$} up to 14\,T. The arrow indicates the direction of increasing magnetic field. The inset shows the linear variation of the magnetization below 1\,T. Typical field dependence of isothermal magnetization at various temperatures for (b) \textit{\textbf{H}}  $\parallel$ \textit{\textbf{a}$_{\rm o}^{\star}$}, (c) \textit{\textbf{H}}  $\parallel$ \textit{\textbf{c}$_{\rm o}^{\star}$} and (d) \textit{\textbf{H}}  $\parallel$ \textit{\textbf{b}$_{\rm o}$}. A field-driven spin-flop transition at $\sim$5.5\,T is observed below $T_{\rm N}$ for \textit{\textbf{H}}  $\parallel$ \textit{\textbf{a}$_{\rm o}^{\star}$} (blue arrow) and \textit{\textbf{H}}  $\parallel$ \textit{\textbf{c}$_{\rm o}^{\star}$} (red arrow).} 
\end{figure}

Keeping in mind the twinned crystal structure, the magnetization along \textit{\textbf{a}$_{\rm o}^{\star}$} and \textit{\textbf{c}$_{\rm o}^{\star}$} represent a mixture of the crystallographic \textit{\textbf{a}$_{\rm o}$} and \textit{\textbf{c}$_{\rm o}$} axes. A large quantitative difference in the magnetization at 5.5\,T along \textit{\textbf{a}$_{\rm o}^{\star}$} and \textit{\textbf{c}$_{\rm o}^{\star}$} makes it unlikely, that a metamagnetic transition occurs at the same field value in the \textit{\textbf{a}$_{\rm o}$} and \textit{\textbf{c}$_{\rm o}$} directions in a single twin domain. Instead, it appears most likely that the increase in the magnetization corresponds to a spin-flop in the \textit{\textbf{a}$_{\rm o}^i$} easy direction of each twin only.
 
Shown in Fig.\,5(b),\,(c) and (d) are the isothermal magnetization at various temperatures for \textit{\textbf{H}}  $\parallel$ \textit{\textbf{a}$_{\rm o}^{\star}$}, \textit{\textbf{H}}  $\parallel$ \textit{\textbf{c}$_{\rm o}^{\star}$} and \textit{\textbf{H}}  $\parallel$ \textit{\textbf{b}$_{\rm o}$}, respectively. The spin-flop transition in \textit{M}(\textit{H}) for \textit{\textbf{a}$_{\rm o}^{\star}$} and \textit{\textbf{c}$_{\rm o}^{\star}$} shifts to lower fields under increasing temperature and vanishes above $T_{\rm N}$. In contrast, the variation in \textit{M}(\textit{H}) along \textit{\textbf{b}$_{\rm o}$} is essentially temperature independent at and above $T_{\rm N}$.

\begin{figure}
	\includegraphics[width=1.0\linewidth]{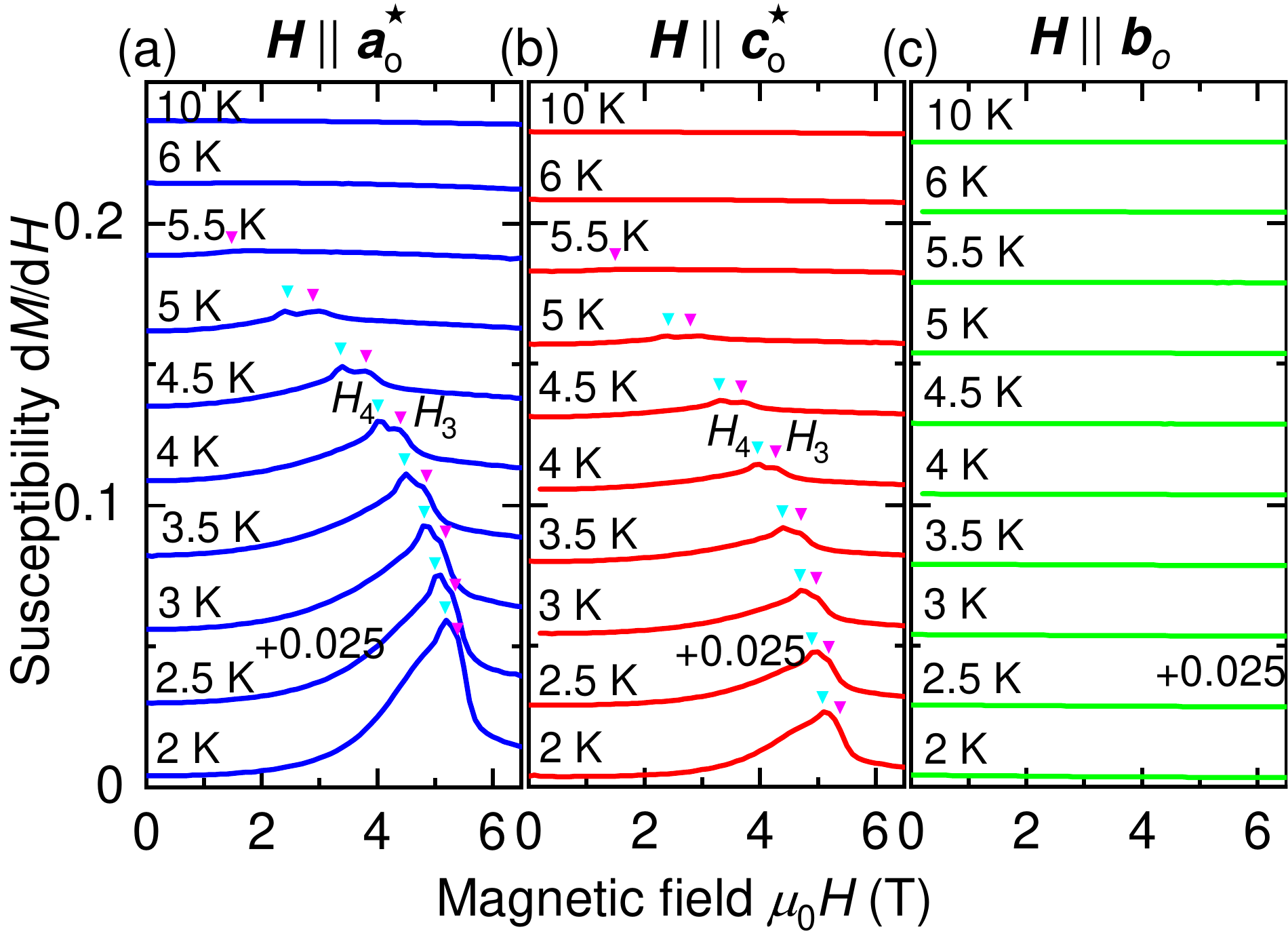}
	\caption{\label{figure6}(a) Susceptibility, d\textit{M}\slash d\textit{H}, calculated from the measured magnetization of orthorhombic CePdAl$_{3}$ for (a) \textit{\textbf{H}}  $\parallel$ \textit{\textbf{a}$_{\rm o}^{\star}$}, (b) \textit{\textbf{H}}  $\parallel$ \textit{\textbf{c}$_{\rm o}^{\star}$} and (c) \textit{\textbf{H}}  $\parallel$ \textit{\textbf{b}$_{\rm o}$}. Data are shifted by 0.025 for clarity. Peaks correspond to field-induced transitions marked by arrows at $H_3$ (pink) and $H_4$ (sky blue). The peaks disappear above $T_{\rm N}$.} 
\end{figure}

In order to trace the field-driven magnetic transition, we have calculated the differential susceptibility, d\textit{M}\slash d\textit{H} from the isothermal magnetization at various temperatures presented in Fig.\,6(a),\,(b), and (c) for \textit{\textbf{H}}  $\parallel$ \textit{\textbf{a}$_{\rm o}^{\star}$}, \textit{\textbf{H}}  $\parallel$ \textit{\textbf{c}$_{\rm o}^{\star}$} and \textit{\textbf{H}}  $\parallel$ \textit{\textbf{b}$_{\rm o}$}, respectively. For fields along \textit{\textbf{a}$_{\rm o}^{\star}$} and \textit{\textbf{c}$_{\rm o}^{\star}$}, the transition is characterized by a broad peak at $\sim$5.2\,T at 2\,K, which resolves into two peaks at elevated temperatures. These peaks exist below $T_{\rm N}$ as marked by arrows at the transition fields $H_3$ and $H_4$, following the labelling scheme described in Sec.\,II. With increasing temperature the field range between the peaks increases and both peaks shift to lower field values. No indication exists of a field-induced transition in d\textit{M}\slash d\textit{H} for field along \textit{\textbf{b}$_{\rm o}$}.

The evolution of the field-induced transitions may be traced in further detail by the temperature dependence of the magnetization \textit{M}(\textit{T}) and the ac susceptibility Re$\chi$$_{\rm ac}$(\textit{T}). Shown in Fig.\,7 is \textit{M}(\textit{T}) and Re$\chi$$_{\rm ac}$(\textit{T}) at various fields up to 14\,T. By decreasing the temperature, orthorhombic CePdAl$_{3}$ undergoes a phase transformation from paramagnetism to antiferromagnetic order at a transition temperature $T_1$ (marked by red lines). This transition is visible in Re$\chi$$_{\rm ac}$(\textit{T}) in all crystallographic directions. The transition at $T_1$ shifts to lower temperatures under increasing field but does not vanish upto the highest field of 14\,T studied. In the intermediate field range from 2\,T to 6\,T, clear changes in \textit{M}(\textit{T}) and Re$\chi$$_{\rm ac}$(\textit{T}) for field along \textit{\textbf{a}$_{\rm o}^{\star}$} [Fig.\,7(a) and (d)]) and \textit{\textbf{c}$_{\rm o}^{\star}$} [Fig.\,7(b) and (e)] point to two additional phase transitions at temperatures denoted $T_3$ (blue line) and $T_4$ (green line). These transitions disappear at fields above 6\,T. For \textit{\textbf{H}}  $\parallel$ \textit{\textbf{b}$_{\rm o}$}, only the first transition at $T_1$ is observed in \textit{M}(\textit{T}) and Re$\chi$$_{\rm ac}$(\textit{T}) [Fig.\,7(c) and (f)].\\ 

\begin{figure}
	\includegraphics[width=1.0\linewidth]{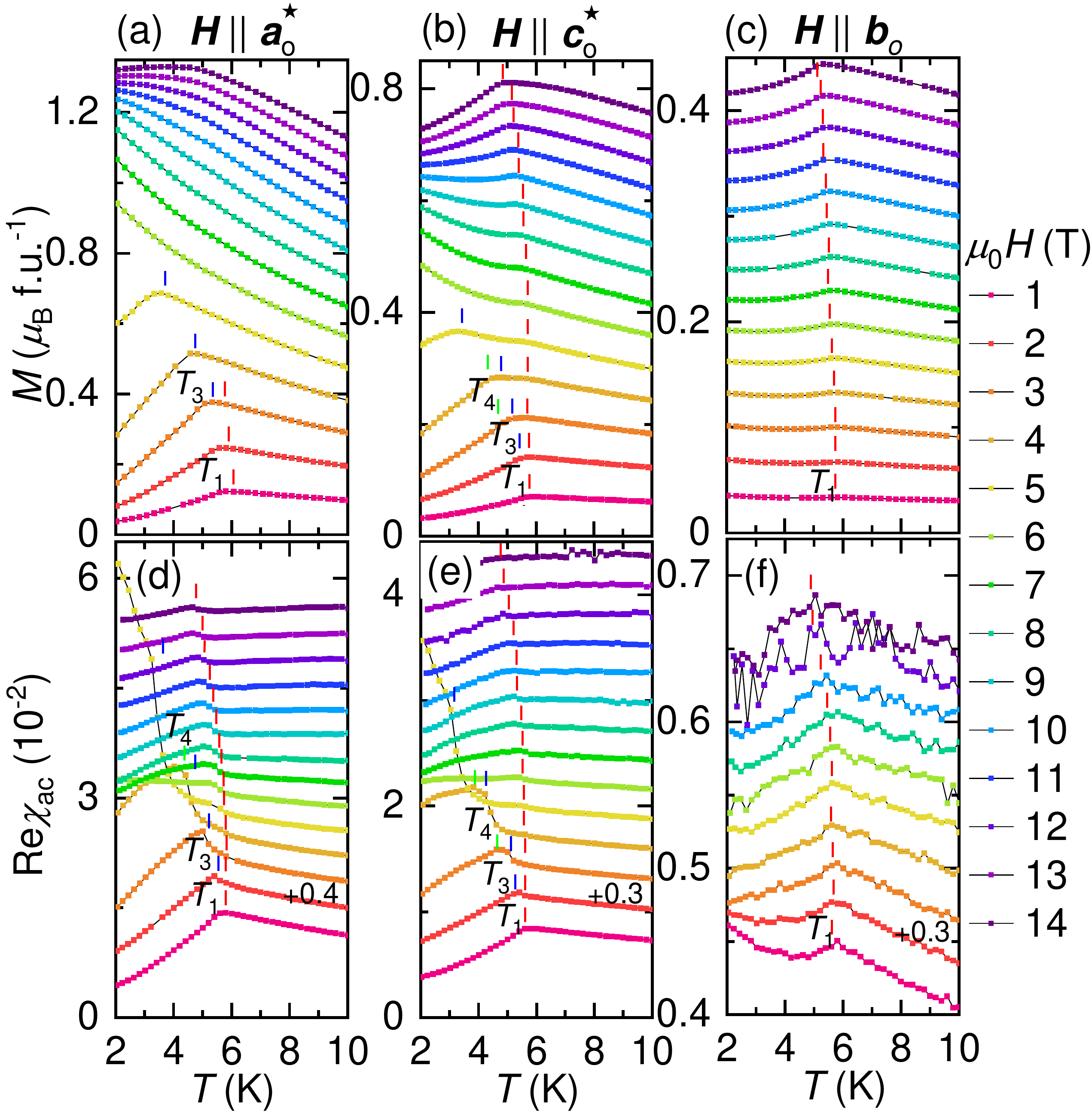}
	\caption{\label{figure7}Temperature dependence of magnetization, \textit{M}(\textit{T}) and real part of ac susceptibility, Re$\chi$$_{\rm ac}$(\textit{T}) of orthorhombic CePdAl$_{3}$ in magnetic fields up to 14\,T. \textit{M}(\textit{T}) is shown in panels (a), (b) and (c), and Re$\chi$$_{\rm ac}$(\textit{T}) in (d), (e) and (f) for \textit{\textbf{H}}  $\parallel$ \textit{\textbf{a}$_{\rm o}^{\star}$}, \textit{\textbf{H}}  $\parallel$ \textit{\textbf{c}$_{\rm o}^{\star}$} and \textit{\textbf{H}}  $\parallel$ \textit{\textbf{b}$_{\rm o}$}, respectively. Re$\chi$$_{\rm ac}$(\textit{T}) is shifted for clarity. Magnetic transitions are marked by vertical lines at temperatures $T_1$ (red), $T_3$ (blue) and $T_4$ (green). A complex behavior with multiple transitions is present for field along \textit{\textbf{a}$_{\rm o}^{\star}$} and \textit{\textbf{c}$_{\rm o}^{\star}$} between 2\,T and 6\,T.} 
\end{figure}

\subsection{Field-dependence of the specific heat}

The specific heat of orthorhombic CePdAl$_{3}$ as a function of temperature at different magnetic fields for \textit{\textbf{H}}  $\parallel$ \textit{\textbf{c}$_{\rm o}^{\star}$} is presented in Fig.\,8. At zero magnetic field [Fig.\,8(b)], a broad shoulder with a point of inflection is observed at $T_1$ followed by a sharp peak at $T_2$. Increasing the applied field results in a broadening of the peak at $T_2$ [Fig.\,8(c)] and a splitting with an additional peak emerging at a lower temperature $T_4$. For even higher fields up to 6\,T [Fig.\,8(d) to (h)], the position of $T_4$ continues to shift to lower temperatures with the emergence of another peak at $T_3$. The emergence of the peaks at $T_3$ and $T_4$ in the specific heat in the intermediate field range from 2 to 6\,T is consistent with the phase transitions deduced from the magnetization and the ac susceptibility (see Figs.\,6 and 7). For fields above 6\,T, a noticeable shift of $T_1$ and $T_2$ to lower temperatures is observed.\\ 

\begin{figure}
	\includegraphics[width=1.0\linewidth]{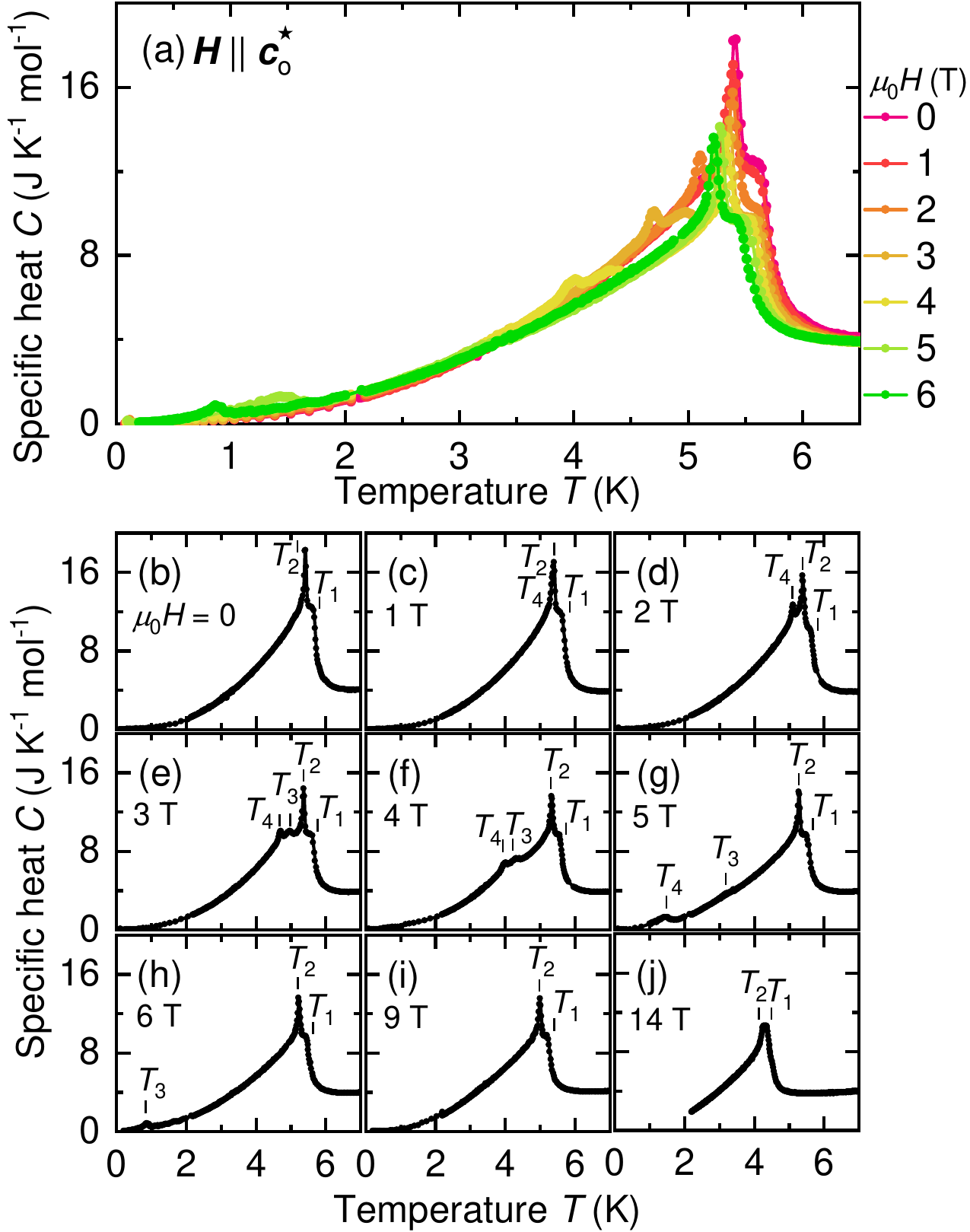}
	\caption{\label{figure8}Specific heat of orthorhombic CePdAl$_{3}$ as a function of temperature under selected magnetic fields up to 14\,T applied along the \textit{\textbf{c}$_{\rm o}^{\star}$} axis. Data measured in the Dryogenic system between 0.08\,K and 4\,K are combined with data measured in the PPMS above 2\,K. At $H=0$ the magnetic transition displays a peak at $T_2$ preceded by a broad shoulder with a point of inflection at $T_1$. Additional peaks emerge at $T_3$  and $T_4$ for magnetic fields between 2\,T and 6\,T.} 
\end{figure}

\subsection{Field-dependence of the magnetic susceptibility}

In order to investigate the qualitative difference between the transitions labelled as $H_3$ and $H_4$ in d\textit{M}\slash d\textit{H} (see Fig.\,6), we have measured the magnetic susceptibility as a function of magnetic field between 0 and 14\,T. Figure 9 shows the real part of the ac susceptibility, Re$\chi$$_{\rm ac}$, and the susceptibility calculated from the magnetization, d\textit{M}\slash d\textit{H}, as a function of increasing and decreasing field.  At 2\,K, d\textit{M}\slash d\textit{H} exhibits two peaks under increasing field, first, a pronounced peak at 5.15\,T, followed by a second broad peak at 5.3\,T for both \textit{\textbf{H}}  $\parallel$ \textit{\textbf{c}$_{\rm o}^{\star}$} [Fig.\,9(a) and (d)] and \textit{\textbf{H}}  $\parallel$ \textit{\textbf{a}$_{\rm o}^{\star}$} [Fig.\,9(c) and (f)]. The first peak shifts to 5\,T resulting in a hysteresis, while the second peak remains at the same field value under decreasing field. Similar effects exists in Re$\chi$$_{\rm ac}$ where the first peak becomes less pronounced with a smaller hysteresis and a slightly lower field of 5.05\,T. Also, the value of Re$\chi$$_{\rm ac}$ is slightly lower around the transition. At higher temperatures, both peaks are shifted to lower field values. The hysteresis in Re$\chi$$_{\rm ac}$ decreases significantly and drops below the noise level at 5\,K [Fig.\,9(b) and (e)]. Here, the magnitude of Re$\chi$$_{\rm ac}$ matches well with d\textit{M}\slash d\textit{H} except around the first peak. The difference in character of the transitions labelled as $H_3$ and $H_4$ suggest their intrinsic origin rather than being related to the twinned microstructure.

On the one hand, the hysteresis observed in d\textit{M}\slash d\textit{H} and Re$\chi$$_{\rm ac}$ is reminiscent of changes of population of multidomain states. On the other hand, the smaller amplitude of Re$\chi$$_{\rm ac}$ as compared to d\textit{M}\slash d\textit{H} indicates the presence of slow relaxation processes around the phase transition. Similar features are known to trace spin textures like helimagnetic disclination or skyrmions in magnetic materials \cite{bauer2012magnetic,bauer2017symmetry,tokunaga2015new}. Further experimental investigations are needed to explore such a possibility in orthorhombic CePdAl$_{3}$.

\begin{figure}
	\includegraphics[width=1.0\linewidth]{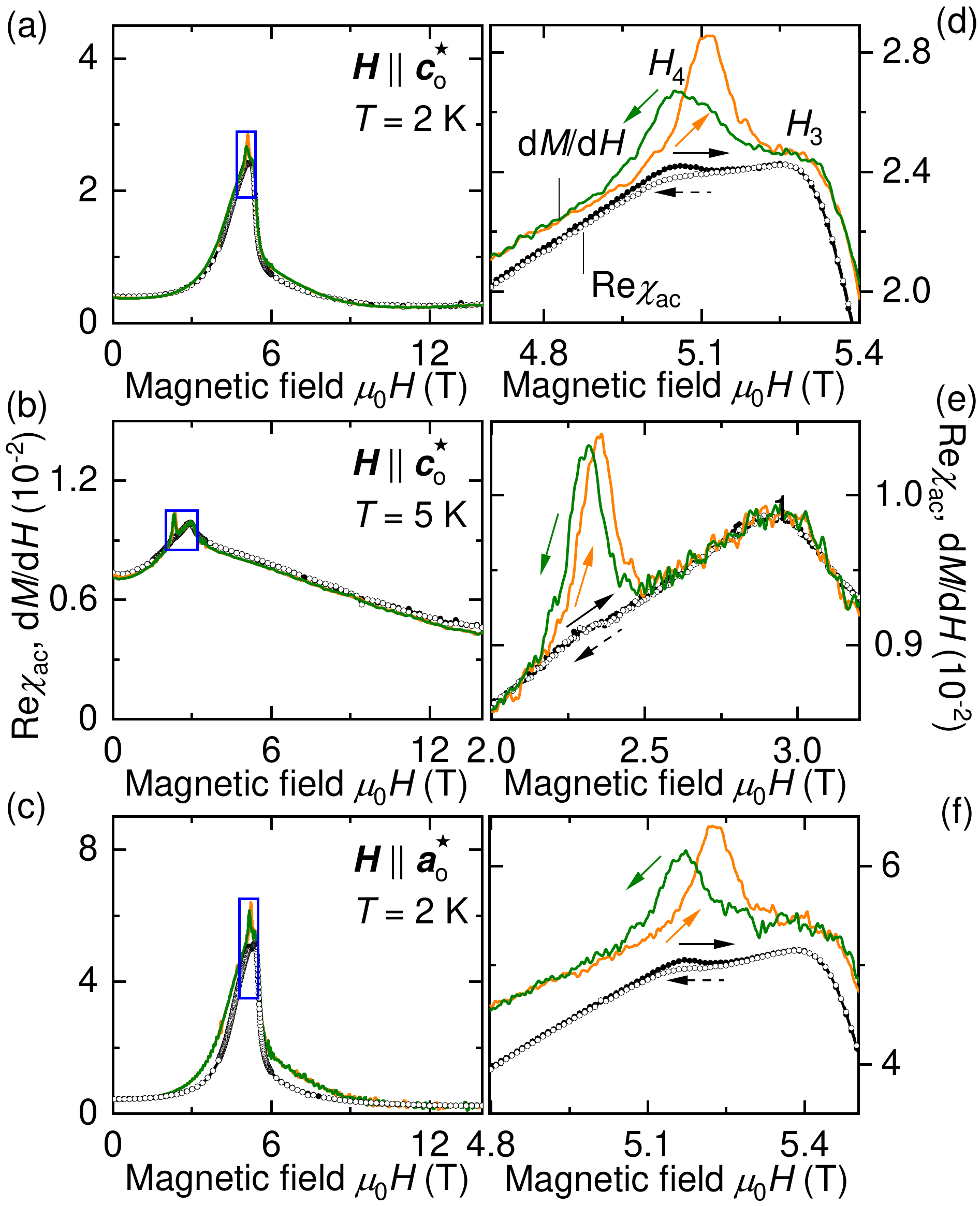}
	\caption{\label{figure10} Details of the magnetic transitions labelled as $H_3$ and $H_4$. Shown are the real part of ac susceptibility, Re$\chi$$_{\rm ac}$, and the susceptibility calculated from the magnetization, d\textit{M}\slash d\textit{H} of orthorhombic CePdAl$_{3}$ as a function of increasing and decreasing field for (a) \textit{\textbf{H}} $\parallel$ \textit{\textbf{c}$_{\rm o}^{\star}$} at 2\,K, (b) \textit{\textbf{H}} $\parallel$ \textit{\textbf{c}$_{\rm o}^{\star}$} at 5\,K, and (c) \textit{\textbf{H}} $\parallel$ \textit{\textbf{a}$_{\rm o}^{\star}$} at 2\,K. (d), (e) and (f) show the magnetic transition regions corresponding to the blue rectangles in (a), (b) and (c), respectively. Colors denote d\textit{M}\slash d\textit{H} for increasing (orange) and decreasing (green) magnetic field. Black circles correspond to Re$\chi$$_{\rm ac}$ for increasing (filled symbols) and decreasing (open symbols) field, respectively. d\textit{M}\slash d\textit{H} was calculated after smoothing the data.} 
\end{figure}
	
\subsection{Magnetic phase diagram}

Combining the features detected in the magnetization and the specific heat, we infer the magnetic phase diagrams for field parallel to \textit{\textbf{c}$_{\rm o}^{\star}$} and \textit{\textbf{b}$_{\rm o}$} shown in Fig.\,10(a) and (b), respectively. Due to the twinned microstructure, the response of the magnetization, specific heat, and ac susceptibility are qualitatively alike for \textit{\textbf{H}}  $\parallel$ \textit{\textbf{a}$_{\rm o}^{\star}$} and \textit{\textbf{H}}  $\parallel$ \textit{\textbf{c}$_{\rm o}^{\star}$}. In addition, the enhanced signal observed along \textit{\textbf{a}$_{\rm o}^{\star}$} as compared to \textit{\textbf{c}$_{\rm o}^{\star}$} indicates that \textit{\textbf{a}$_{\rm o}^{\star}$} reflects a larger fraction of the easy axis, \textit{\textbf{a}$_{\rm o}$}. Therefore, the transitions along both \textit{\textbf{a}$_{\rm o}^{\star}$} and \textit{\textbf{c}$_{\rm o}^{\star}$} reflect equally the phenomenon belonging to the easy \textit{\textbf{a}$_{\rm o}$} axis of the untwinned single domain.

Four magnetic regions may be distinguished for field along \textit{\textbf{c}$_{\rm o}^{\star}$}, denoted AF-I, AF-II, AF-III and AF-IV. At low temperature and zero-field, the ground state is denoted as AF-I. With increase temperature, AF-II appears at 5.4\,K before entering in the paramagnetic (PM) state above 5.6\,K. Signatures of the AF-II region are detected only in the specific heat. The application of a magnetic field at low temperature drives a spin-flop transition from AF-I to AF-IV with an intermediate region AF-III in a narrow field range only. For finite field applied along the hard axis, i.e., \textit{\textbf{H}}  $\parallel$ \textit{\textbf{b}$_{\rm o}$} [Fig.\,10(b)] only the AF-I and PM phases were observed, possibly due to the lack of specific heat data for finite fields along the hard axis. However, the AF-II transition was observed in zero field and the AF-II regime is shown in the phase diagram in Fig.\,10(b) for consistency. 

While the magnetization suggests a collinear antiferromagnetic structure along \textit{\textbf{a}$_{\rm o}$} in the AF-I phase, and AF-IV shows the characterisics of a spin-flop phase, the nature of AF-II and AF-III remain completely unknown. Neutron scattering studies under magnetic field are needed to determine the nature of the four antiferromagnetic phases we observed in orthorhombic single crystal CePdAl$_{3}$.

\begin{figure}
	\includegraphics[width=1.0\linewidth]{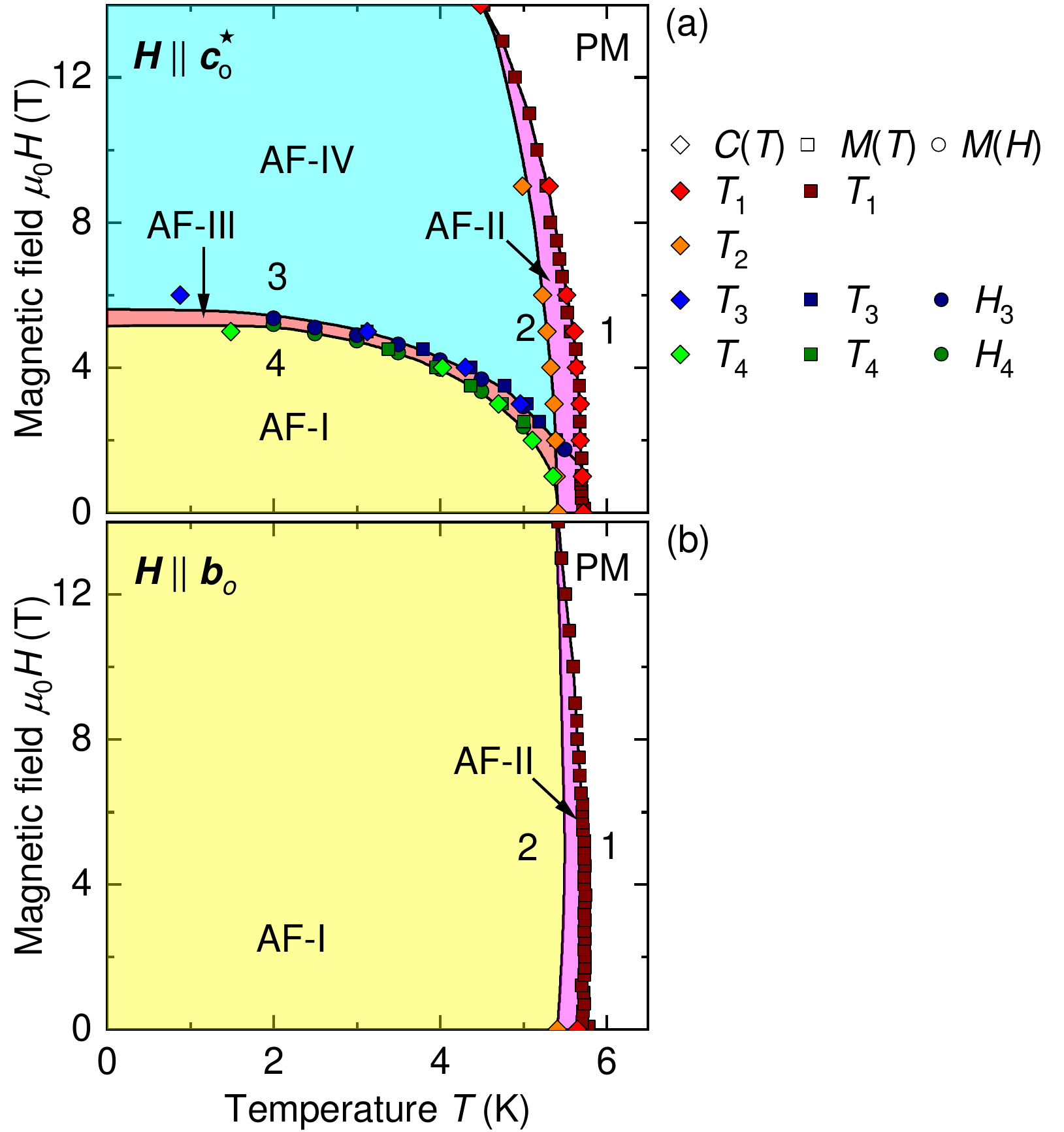}
	\caption{\label{figure9} Magnetic phase diagram of orthorhombic CePdAl$_{3}$ for (a) \textit{\textbf{H}}  $\parallel$ \textit{\textbf{c}$_{\rm o}^{\star}$} and (b) \textit{\textbf{H}}  $\parallel$ \textit{\textbf{b}$_{\rm o}$} as inferred from the magnetization and specific heat. Due to crystal twinning, the phase diagram for \textit{\textbf{H}}  $\parallel$ \textit{\textbf{a}$_{\rm o}^{\star}$} qualitatively resembles the phase diagram for \textit{\textbf{H}} $\parallel$ \textit{\textbf{c}$_{\rm o}^{\star}$} shown in (a). Phase transitions are guided by the lines which are denoted by numerals $j$\,=\,1, 2, 3, and 4. The associated temperature and field values are labelled as $T_j$ and $H_j$, respectively. Four magnetically ordered phases may be distinguished as discussed in the text.} 
\end{figure} 
	
\section{Conclusions}

In summary, we measured the magnetization, ac susceptibility, and specific heat of a single crystal of CePdAl$_{3}$ grown by optical float-zoning. A highly anisotropic behavior with a twinned orthorhombic crystal symmetry was observed. Antiferromagnetic order with $T_{\rm N}=5.6$\,K was observed in terms of transitions in the ac susceptibility and specific heat. The magnetization is characteristic of antiferromagnetic order with an easy \textit{\textbf{a}$_{\rm o}$} direction in the basal plane. Field-driven transitions were detected in the magnetization along the easy direction, consistent with the ac susceptibility and specific heat. Taken together, our study reveals a strong interplay of electronic correlations, complex magnetic order and structural modifications in CePdAl$_{3}$.

\begin{acknowledgments}

We wish to thank A.\ Engelhardt, S.\ Mayr, and W.\ Simeth for fruitful discussions and assistance with the experiments. We thank T. E. Schrader on measurements with the Rigaku single-crystal diffractometer in the x-ray labs of the Jülich Centre for Neutron Science (JCNS). This study has been funded by the Deutsche Forschungsgemeinschaft (DFG, German Research Foundation) under TRR80 (From Electronic Correlations to Functionality, Project No.\ 107745057, Project E1), SPP2137 (Skyrmionics, Project No.\ 403191981, Grant PF393/19), and the excellence cluster MCQST under Germany's Excellence Strategy EXC-2111 (Project No.\ 390814868). Financial support by the European Research Council (ERC) through Advanced Grants No.\ 291079 (TOPFIT) and No.\ 788031 (ExQuiSid) is gratefully acknowledged.

\end{acknowledgments}

\bibliography{ref}
		
\end{document}